\documentclass[prl,nobibnotes,floatfix,superscriptaddress,twocolumn]{revtex4-1}
\usepackage{amsfonts,amsmath,graphicx,epsfig,latexsym}
\usepackage{subfigure}
\usepackage{graphicx}
\usepackage{color}
\usepackage{natbib}
\usepackage{multirow}
\usepackage{latexsym,amssymb}
\usepackage{amsmath}
\usepackage{soul}
\usepackage{ulem}
\usepackage[linktocpage,bookmarksopen,bookmarksnumbered]{hyperref}

\begin{document}
\title{ Systematic beyond-DFT study of binary transition metal oxides}
\author{Subhasish Mandal}
\affiliation
	{Department of Physics and Astronomy, Rutgers University, Piscataway,  USA}

\author{Kristjan Haule}
\affiliation
{Department of Physics and Astronomy, Rutgers University, Piscataway,  USA}

\author{Karin M. Rabe}
\affiliation
{Department of Physics and Astronomy, Rutgers University, Piscataway,  USA}

\author{ David Vanderbilt}
\affiliation
{Department of Physics and Astronomy, Rutgers University, Piscataway,  USA}

\begin{abstract}
{\footnotesize

Various methods going beyond density-functional theory (DFT), such as DFT+U, hybrid functionals, meta-GGAs, GW, and DFT-embedded dynamical mean field theory (eDMFT), have been developed to describe the electronic structure of correlated materials, but it is unclear how accurate these methods can be expected to be when applied to a given strongly correlated solid. It is thus of pressing interest to compare their accuracy as they apply to different categories of materials. Here, we introduce a novel paradigm in which a chosen set of beyond-DFT methods is systematically and uniformly tested on a chosen class of materials.  For a first application, we choose the target materials to be the binary transition-metal oxides FeO, CoO, MnO, and NiO in their antiferromagnetic phase and present a head-to-head comparison of spectral properties as computed using the various methods. We also compare with available experimental angle-resolved photoemission spectroscopy (ARPES), inverse-photoemission spectroscopy, and with optical absorption. We find both B3LYP and eDMFT can reproduce the experiments quite well, with eDMFT doing best in particular when comparing with the ARPES data. 
}
\end{abstract}

\maketitle
\newpage

\section{Introduction}

Future technologies depend on new materials with tailored, enhanced and/or novel functionalities. Strongly correlated materials exhibit rich physics that offers unique opportunities in this regard, particularly in their magnetic, optical and transport properties~\cite{doi:10.1063/1.1712502}. While progress in computational materials design has greatly accelerated the process of identifying and realizing materials with moderate correlations, significant challenges remain for the quantitative, and in some cases even qualitative, computational prediction of the properties of interest of strongly correlated materials. Furthermore, accurate computation of excited-state properties even in moderately correlated materials is beyond the scope of density functional theory (DFT) ~\cite{onida_electronic_2002,DMFT2,DMFT3}, and in strongly correlated materials, it is well known that DFT can even fail correctly to predict whether a system is a metal or an insulator. The existing materials databases~\cite{Mdata}, constructed in the spirit of the Materials Genome Initiative~\cite{MGI}, are built almost exclusively by DFT engines and are thus very often making
incorrect predictions in correlated materials. There is thus a longstanding interest in developing perturbative, stochastic, and hybrid-functional approaches, referred to as ``beyond-DFT" methods, to treat strongly correlated and excited-state properties.

In strongly correlated materials, one reason for the failure of DFT is the delocalization or self-interaction error~\cite{PZ,RevModPhys.61.689}, which can be partially fixed by adding a Hubbard $U$ in the DFT+U approach~\cite{Anisimov_1997}. This method recovers the insulating state in many materials that are incorrectly predicted to be metallic in DFT~\cite{Anisimov_1997,doi:10.1002/qua.24521,Gopal_2017}. 
Moreover, DFT and DFT+U methods give quite accurate results for structural parameters in most materials~\cite{RevModPhys.64.1045,Mdata}.
However, unique determination of an appropriate value of $U$ for more general quantitative calculations has proved surprisingly problematic.
Hybrid functionals~\cite{PhysRevB.74.155108,doi:10.1063/1.1564060}, which also correct most of the self-interaction error by incorporating a certain fraction of exact exchange, significantly improve the descriptions of many {\it d}-electron systems~\cite{PhysRevB.74.155108,doi:10.1063/1.1564060,PBE0,LYP}. 
The fraction of exact exchange is treated as a tuning parameter, generally falling in the range $\sim$ 0.2 to 0.45. 

Many-body perturbation theory in the GW quasiparticle approximation~\cite{hedin_new_1965} is a beyond-DFT method that was developed to better describe the quasiparticle excitations in solids, and results compare well with the experimental PES/IPES for many semiconducting and insulating systems with open {\it s} and {\it p}-shells~\cite{HL, aryasetiawan_gw_1998,onida_electronic_2002}.
Dynamical mean-field theory (DMFT)~\cite{DMFT2} gives an exact treatment of intrashell correlations local to a given ion. Originally developed for lattice models, its combination with DFT in the so-called DFT+DMFT method~\cite{DMFT3} allows for a quantitative description of the electronic structure of strongly correlated materials such as Fe-based superconductors~\cite{haule3,PhysRevLett.119.067004,haule_spin,PhysRevB.90.060501,PhysRevB.90.060501,PhysRevB.92.195128}, Mott insulators~\cite{Kunes:2008bh}, and heavy fermion systems ~\cite{Shim} without tuning parameters~\cite{covalency}~\cite{Kent348}.
Furthermore, DFT+DMFT calculations of the spectral function have been instrumental for the understanding of photoemission and inverse photoemission spectroscopy (PES/IPES) for a variety of strongly correlated systems.

 \begin{figure*}
\includegraphics[width=500pt, angle=0]{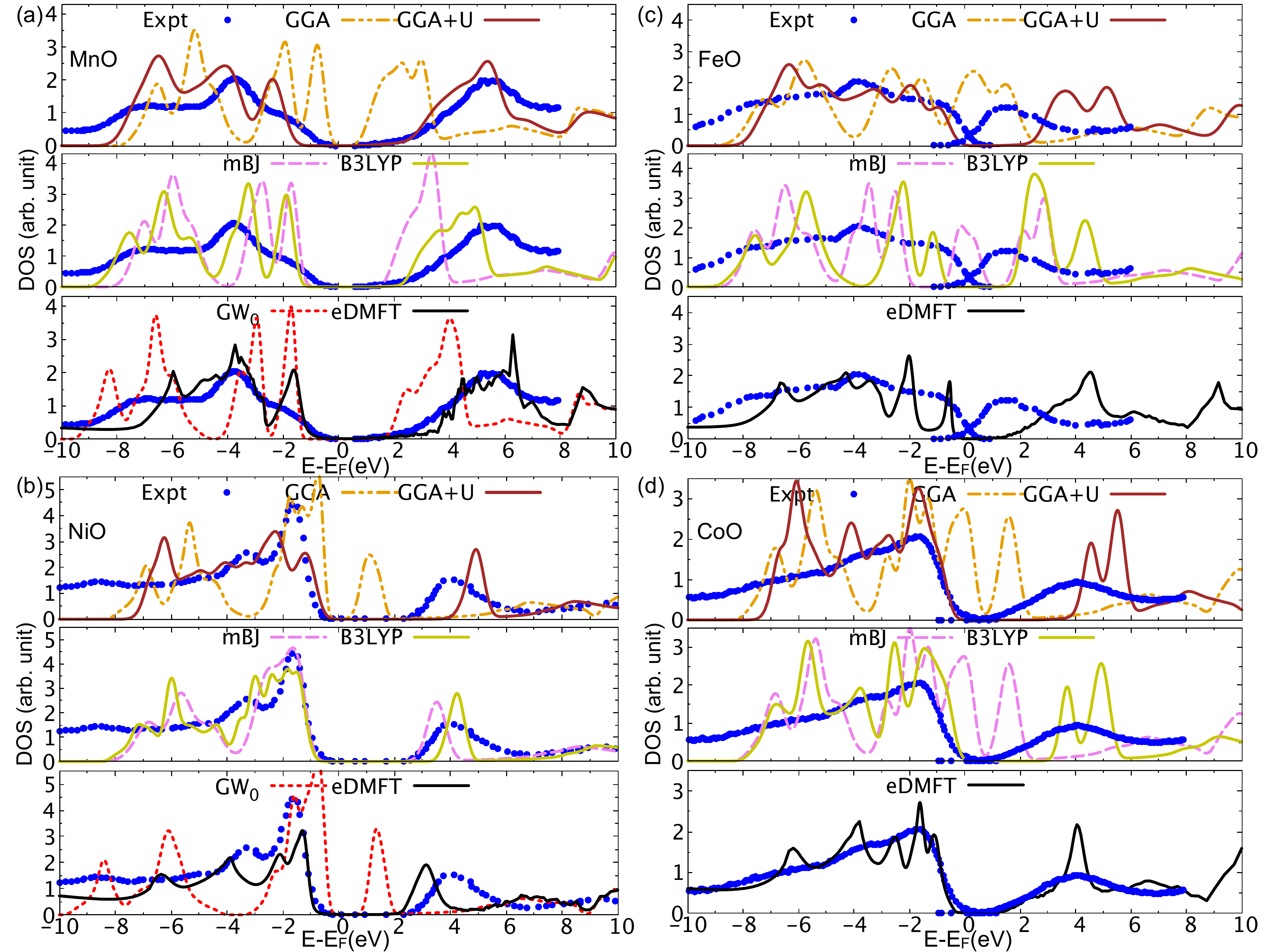}
\caption{ (Color online)
Total density of states (states/eV) as computed in GGA, GGA+U, mBJ, B3LYP, GW$_0$, and  eDMFT for (a) MnO, (b)NiO, (c) FeO, and (d)CoO. Blue dots indicate photoemission and inverse photoemission data in arbitrary unit as obtained form  Ref~\cite{PhysRevB.44.1530,PhysRevLett.53.2339,PhysRevB.44.6090,Zimmermann_1999} for MnO, NiO, CoO, and FeO respectively.
}
\end{figure*}

The beyond-DFT methods vary greatly in their suitability for different classes of correlated materials. In addition, they are considerably more computationally intensive than DFT or DFT+U. Especially in the context of high-throughput studies, this means that there is a pressing need for a systematic way to choose, for any given material, the computational method that will give physically accurate results without unnecessary computation. Development of this capability requires that the performance of various beyond-DFT methods be systematically and uniformly tested on a diverse training set of strongly correlated materials that are experimentally well characterized. 
 
Binary transition metal oxides (TMO) are among the most thoroughly studied strongly correlated materials~\cite{Anisimov_1997,doi:10.1002/qua.24521,Gopal_2017,PhysRevB.82.045108,
PhysRevLett.100.066406,PhysRevB.96.045111,PhysRevB.92.115118,PhysRevB.74.155108,PhysRevB.54.13566,Cohen31011997,PhysRevLett.108.026403,PhysRevLett.99.156404,PhysRevB.44.3604,PhysRevB.44.1530,PhysRevLett.53.2339,PhysRevB.44.6090,Zimmermann_1999}, and thus are a natural starting point for generation of the training set. They include a number of wide-gap insulators predicted to be metallic in the conventional density functional formalism (DFT)~\cite{PhysRevB.30.4734,Gopal_2017}. Those that contain early and late transition metals are usually categorized as ``Mott" and ``charge-transfer" insulators in the ``Zaanen-Sawatzky-Allen" scheme~\cite{PhysRevLett.55.418}, and are insulating both above and below the Neel ordering temperature, with strongly localized $3d$ magnetic moments.  These localized magnetic moments, originating in the $3d$ TM states, hybridize with more itinerant $4s$ and oxygen $2p$ states, resulting in competition between localization and itinerancy~\cite{PhysRevLett.100.066406,PhysRevB.96.045111}. Another reason that binary TMOs are ideal for the training set is that their crystal structures are very simple~\cite{PhysRevB.74.155108}. In the paramagnetic phase, they crystallize in the rock-salt (Fm3m) structure; at lower temperatures, antiferromagnetic ordering (AFM II)~\cite{PhysRev.110.1333} results in a rhombohedral (R3m) structure, with two transition metal ions in the unit cell. 

In this paper, we systematically and uniformly test various beyond-DFT methods on the set of binary transition metal oxides MnO, FeO, CoO and NiO, allowing a head-to-head comparison between the various methods with experimental photoemission and inverse photoemission measurements. The methods included in the study are GGA with the PBE functional~\cite{PBE}, GGA+U with the Anisimov and Lichtenstein formalism~\cite{PhysRevB.48.16929,PhysRevB.52.R5467}, meta-GGA with the modified Becke-Johnson (mBJ) potential~\cite{mbj}, all-electron GW$_0$ ~\cite{FHI} in the Hedin formalism, hybrid functionals with B3LYP ~\cite{becke_new_1993}, and DFT+DMFT in the DFT+Embedded DMFT (eDMFT) formalism ~\cite{eDMFT2010,eDMFT2018,edmft}. In addition, optical properties are computed with B3LYP and eDMFT and compared with available experiments. We expect to expand the training set of materials dramatically in future work, with the eventual goal of constructing a database in which a chosen set of DFT and beyond-DFT methods are systematically applied to an increasingly wide range of materials.  Such a database holds great promise to enhance the power of computational materials design and discovery.

 The calculations are performed for MnO, FeO, CoO and NiO in the rocksalt structure~\cite{PhysRevB.74.155108} with experimentally reported lattice parameters a=4.445~$\textrm{\AA}$~\cite{MnO_expt}, 4.334~$\textrm{\AA}$~\cite{FeO_expt}. Magnetic ordering is taken as AFM II along [111]~\cite{PhysRev.110.1333}, which leads to a rhombohedral (R3m) space group-symmetry with two transition-metal ions in the unit cell. Calculations for the paramagnetic state will be considered separately in a future publication (Ref.~\cite{future_SM}).

\section{Results}

\begin{figure*}
\includegraphics[width=440pt, angle=0]{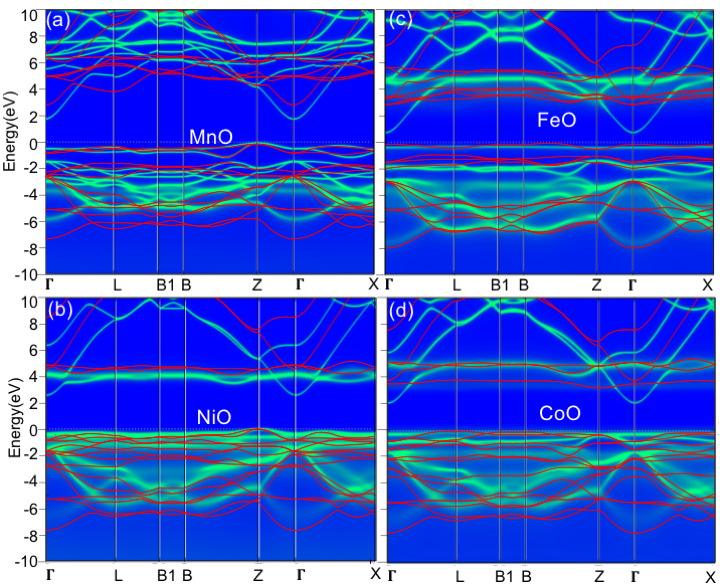}
\caption{ (Color online)
Band structure and spectral function as computed in B3LYP (in red) and eDMFT  (green) for (a)MnO, (b)NiO, (c)FeO, and (d)CoO. 
}
\end{figure*}

\subsection{Density of states}
The density of states (DOS) is shown in Fig.~1 as obtained by GGA, GGA+U, mBJ, GW, B3LYP, and eDMFT methods. From the computed density of states, we see directly whether a material is predicted to be a metal or insulator with a given method. Another important point of comparison is the splitting between the peak at the top of the valence band and the lowest peak in the conduction band, which is quantified experimentally as the PES/IPES gap. For comparison, in each subplot, we include the experimental photoemission (PES) and inverse-photoemission (IPES) for MnO, NiO, CoO, and FeO from Ref.~\cite{PhysRevB.44.1530,PhysRevLett.53.2339,PhysRevB.44.6090,Zimmermann_1999}. Since the experimental PES and IPES spectra are reported in arbitrary units, we have arbitrarily re-scaled them to fit in the range of the computed DOS. Note also that we compare them to the total DOS, which is the best choice for off-resonance spectra of PES/IPES, and we expect to see correct peak positions and similar intensities, although the precise intensity of each peak is not expected to be achieved here, as that would require one to compute the matrix elements for PES/IPES processes. Finally, let us mention that experimentally PES is expected to be more precise than IPES, and the latter generally has larger broadening due to experimental resolution.
 
 For MnO, all six methods predict that the AFM systems are insulating. As expected, GGA substantially underestimates the experimental PES/IPES gap. With the value of $U=6.04$~eV~\cite{PhysRevB.74.155108}, GGA+U predicts the experimental PES/IPES gap accurately. Similarly, the mBJ, B3LYP, and GW$_0$ methods also show improved agreement for the MnO peak splitting relative to GGA, without using any material-specific tuning parameters. While it is slightly underestimated by B3LYP, the agreement with the experimental PES/IPES of MnO is best with the eDMFT method.

 For NiO, we find similar results. All methods predict an insulating character, with GGA underestimating the experimental PES/IPES gap. GW$_0$ does not improve the spectrum much from its GGA shape. On the other hand, the meta-GGA with the mBJ functional performs remarkably well. As was the case for MnO, both B3LYP and eDMFT predict the experimental PES/IPES spectra very well (Fig. 1(b)), with a slightly better match in B3LYP. 

FeO and CoO are the most challenging cases, as the regular GGA predicts them to be metallic. Using $U$=5.91 and 6.88 eV for FeO and CoO respectively~\cite{PhysRevB.74.155108}, both the systems recover insulating phase.  mBJ predicts metallic solutions for FeO and CoO, similar to GGA (Fig.1c and 1d). GW$_0$ on top of LDA also predicts both FeO and CoO to be metallic (not shown). Both B3LYP and eDMFT show a very good agreement with the PES/IPES (Fig.~1) for all four TMOs, except for FeO. As was emphasized in the literature~\cite{FeO_nonstochiometric}, FeO crystals tend to be non-stoichiometric, which is the likely cause of inadequate IPES spectrum. It would be desirable to repeat this experiment on more stoichiometric single crystals.
Overall, eDMFT performs best for describing the peak positions. In particular, for MnO and CoO it predicts the gradual increase of the conduction band intensity for the unoccupied states in good agreement with IPES.

\subsection{Spectral function} 
From the above discussion, it is clear that only B3LYP and eDMFT can consistently reproduce the experimental PES/IPES peak positions without using a material-specific tuning parameter for all four TMOs. Hence, we discuss spectral functions as computed by these two methods, shown in Fig.2, which reproduce and go beyond the discussion of the density of states. For example, we see again that
splitting between occupied and unoccupied flat 3{\it d}- bands (previously discussed for the DOS) is larger in eDMFT than in B3LYP for MnO, CoO and FeO, while it is slightly smaller in NiO.
From the spectral function, we also learn about the dispersion of the key bands near the Fermi level. 
In general, we expect B3LYP to show more bands than eDMFT, as all spectral weight in static theories, like B3LYP, needs to come from sharp band excitations, while in eDMFT, part of the spectrum is incoherent and is redistributed as diffuse weight over a large energy range. We showed before (Fig.~1) that except for NiO, eDMFT peak positions in the density of states are in slightly better agreement with PES/IPES, hence we expect the spectral functions of eDMFT are likely a better prediction for angle-resolved inverse photoemission as well. We compare eDMFT spectral functions (green) and B3LYP bandstructure (white) for NiO with the available ARPES, measured by Shen {\it et al.}~\cite{PhysRevB.44.3604} (Fig.3). The ARPES data in NiO were taken in the AFM phase, consistent with our theoretical calculation. The vertical energy axes are shifted to match the experimental bands at the $\Gamma$ point near $\sim-$2 eV, as the position of the chemical potential inside the gap is arbitrary in theory, and determined by the impurity concentration in the experiment.

Within eDMFT, the $4s$ dispersing state appears as the first conduction band at the $\Gamma$ point in all four TMOs and gives a direct gap to be at the $\Gamma$ point. In B3LYP this state similarly appears in MnO as the first conduction excitation but is shifted much higher in all other TMOs. 
The precise position of this state is hard to determine from currently available experimental data. As eDMFT does not treat this $4s$ state as correlated, its position is not improved from its LDA description, and it might be too low in some TMOs. However, IPES does show a very gradual increase in the intensity in MnO and CoO, due to the presence of this $4s$ state, which agrees with eDMFT. Note that in NiO this $4s$ state is shifted upwards compared to the flat $3d$ states in eDMFT, and experimentally IPES in NiO does show a more abrupt and narrower unoccupied peak (Fig.1). It is therefore tempting to speculate that the first conduction state in all four TMOs is such a dispersive $4s$ band at the $\Gamma$ point, as predicted by eDMFT. 

For FeO and CoO, another prominent feature of the eDMFT spectral function is a very flat band just below the Fermi level. This flat band gives rise to a sharp peak in the occupied DOS near E$_F$. Such a flat band is also observed in a GW computation done on top of hybrid functional~\cite{PhysRevB.86.235122}, but is not found in regular DFT+U or hybrid approaches. We notice the similar flat band in the experimental APRES spectrum for CoO in the PM phase ~\cite{future_SM}. For FeO we are not aware of any experimental ARPES data to compare with.

For MnO, the eDMFT spectral function is much sharper than in other TMOs. This is due to the fact that the entire fluctuating moment orders in MnO, which makes the system more mean-field-like and less correlated. Unfortunately, there are no experimental data available for MnO to compare with.

In Fig.~3, we compare the computed spectral functions with experimental ARPES data for NiO, which are the only one of our four materials for which such data are available in the AFM phase. We see that the experimental ARPES data match quite well with the eDMFT predicted spectral function, while the B3LYP bands do not show many similarities with the experiment. In particular, at the $\Gamma$ point, experiment resolves only two peaks, with one flat state around $-2\,$eV and another degenerate state around $-3\,$eV, in which two branches disperse downward, and one remains mostly flat towards the $X$ point. This is all in agreement with eDMFT, while B3LYP shows a very different pattern of degeneracy at the $\Gamma$ point, not matching ARPES. Moreover, B3LYP predicts several extra bands near the $\Gamma$ point, in particular near $-6.5\,$eV, which did not show up in ARPES and are absent in eDMFT. An additional extremely weak spectrum was observed for the uppermost valence band (not shown here), which was only noticed for selected photon energies and certain emission angles in the ARPES experiment~\cite{PhysRevLett.99.156404,PhysRevB.44.3604}. The uncertainties of this spectrum were discussed in Ref.~\cite{PhysRevLett.99.156404,PhysRevB.44.3604}. Finally, for an even better match of eDMFT spectrum and ARPES, one would need to shift the dispersive oxygen $p$ states slightly lower compared to the flat, mostly $3d$ state at $-$2eV. This deficiency of LDA+eDMFT is known and is inherited from LDA, which tends to place oxygen $2p$ states slightly close to the Fermi level.

\begin{figure}
\includegraphics[width=240pt, angle=0]{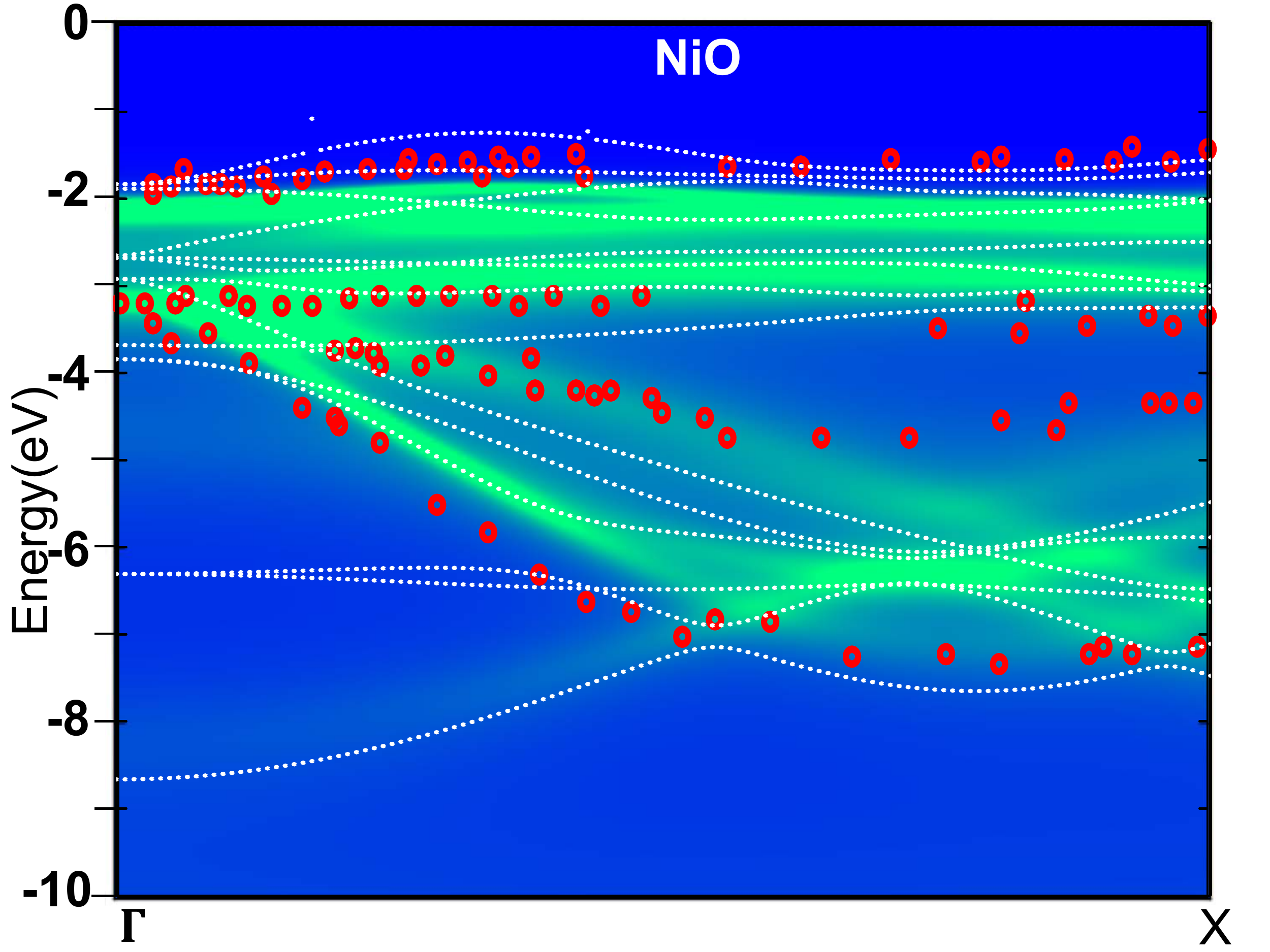}
\caption{ (Color online)
Comparison of experimental ARPES (in red dots) by Shen {\it et al.} with spectral function as computed with eDMFT (green) and band structure with B3LYP (in white lines) for NiO. Experimental data are reproduced from Ref.~\onlinecite{PhysRevB.44.3604}.
}
\end{figure}

\begin{figure*}
\includegraphics[width=460pt, angle=0]{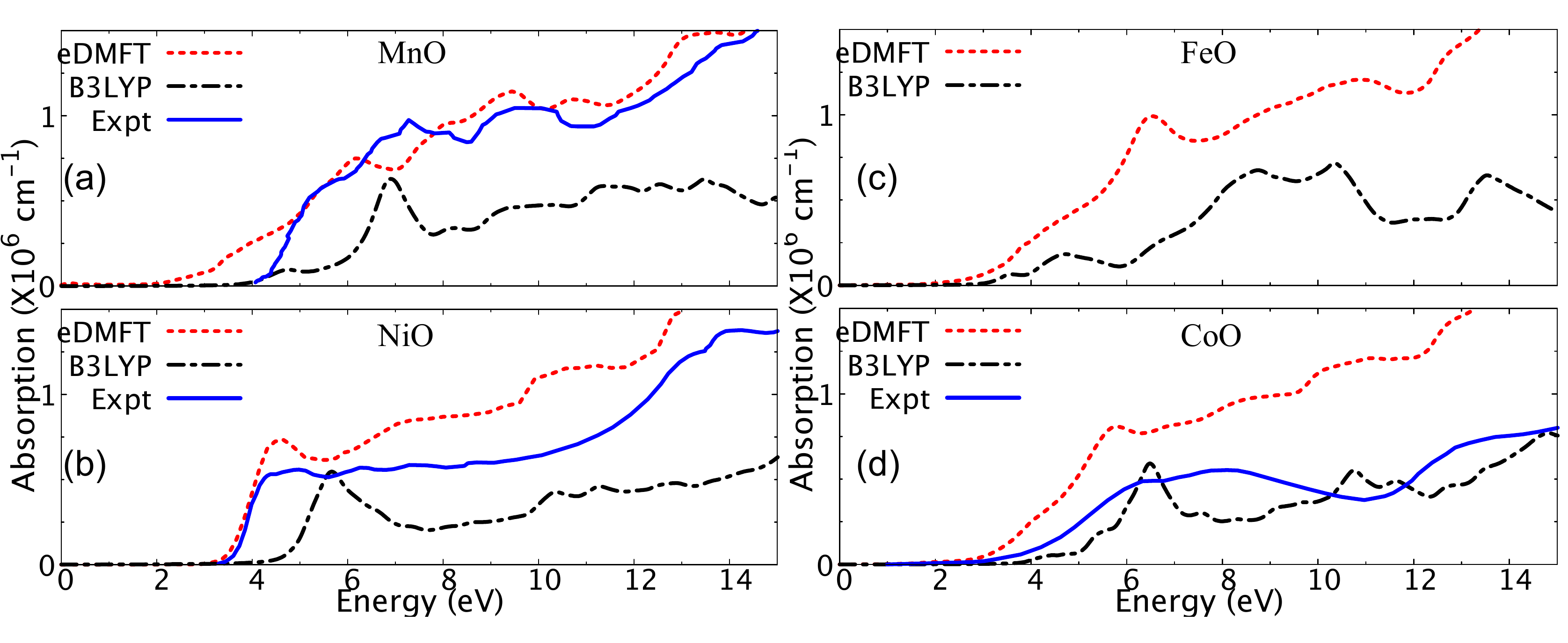}
\caption{ (Color online)
Comparison of eDMFT and B3LYP computed optical absorption coefficients (in dashed lines) for all four TMOs with available experiments (in solid lines). Experimental data are directly obtained from Powell {\it et al.}~\cite{PhysRevB.2.2182} for CoO and NiO and from  R\"odl {\it et al.}~\cite{PhysRevB.86.235122} for MnO, which was used to compare with the experimental reflectivity by Ksendzov {\it et al.}~\cite{MnO-optics}. 
}
\end{figure*}

\subsection{Optical Absorption} 
Finally, we present our results for the optical absorption, which measures the vertical (zero momentum transfer) transitions between the single-particle states, computed by B3LYP and eDMFT.  In Fig.~4 we compare the absorption coefficient from the reflectivity measurements. For MnO, the optical absorption is extracted from figure in R\"old~\textit{et al.}~\cite{PhysRevB.86.235122}, where the measurements by Ksendzov~\textit{et al.}~\cite{MnO-optics} were reproduced. The original data for MnO are not currently accessible. For NiO and CoO, the experimental absorption coefficients are extracted from Powell~\textit{et.al.}~\cite{PhysRevB.2.2182}, which were obtained from the measured reflectively spectra. Reliable reflectivity spectra performed on stoichiometric FeO are unavailable.

We observe various important points in Fig.~4. First, within eDMFT and B3LYP, the onset of the optical absoprtion is gradual in MnO, FeO, and CoO, while it is much more sudden in NiO. This gradual onset in eDMFT is due to the presence of a dispersive $4s$ band around the $\Gamma$ point as discussed above.

Second, the peak positions in the experimental absorption and those computed by eDMFT agree very well. For example, the first peak for NiO in eDMFT is at 4.6 eV while in the experiment it is at 4.9 eV. The overall shapes and magnitudes match very well in NiO. Similarly, for MnO, the overall shapes of eDMFT and experiment match very well. However, the first shoulder of intensity between 2.5-4eV seems to be missing in the data of Ksendzov~\textit{et al.}~\cite{MnO-optics}, which seems somewhat inconsistent with the IPES in Fig.~1, where a slow increase of the intensity is noticed. To clarify the correct placement of the Mn $4s$ states, it would be desirable to acquire new spectra for MnO obtained with modern techniques. 

For CoO, the eDMFT peak positions are similar to the experiment, but the overall match is not so good. B3LYP, which does not match well in MnO and NiO (the intensity is too small and peaks do not align) seems superior in CoO, with an overall good match and a correct gap.

\section{Discussion}

 We find that only B3LYP and eDMFT can properly reproduce the experimental insulating state for all four compounds in the family without artificial tuning of compound-specific parameters. B3LYP still slightly underestimates the experimental peak positions for PES and IPES and the insulating gap for MnO and CoO. eDMFT slightly underestimates the PES/IPES gap in NiO, but overall agrees with the experimental PES/IPES peak positions very well. When comparing with available ARPES, eDMFT compares much better than B3LYP. Many additional high-energy bands are observed in B3LYP that are not present either in eDMFT or identified in the experiment. Computed optical absorptions in NiO and MnO also show better agreement in peak positions and in the total intensity for the absorption coefficient between experiment and eDMFT than B3LYP, but B3LYP performs better in CoO.  As the optical measurements are not recent, we caution that this conclusion might need revision if the optical measurement is redone in CoO.

 The failure of GW$_0$ for these correlated systems is not surprising, and were argued in the literature to be either due to lack of a proper starting point for computing dielectric function or due to the lack of self-consistency in electrons screening~\cite{PhysRevB.82.045108,PhysRevB.79.235114,PhysRevB.86.235122}. The inclusion of Hubbard U in the starting point (applying GW$_0$ on top of LDA+U) also gives an insulating solution for FeO and CoO~\cite{PhysRevB.82.045108,PhysRevB.79.235114,PhysRevB.86.235122}, but this again requires fine-tuning of $U$, like in GGA+U, and is not attempted here.

The good performance of eDMFT is due to the inclusion of higher-order Feynman diagrams by the impurity solver, which allows one to compute the local correlation exactly. It also includes the electron scattering leading to a finite lifetime, which is beyond the limit of hybrid-functional approaches, which only correct the exchange part of the interaction.

We expect that our results for the AFM phase of insulating binary TMOs should be representative of a broader class of moderately correlated materials with open 3$d$ shells and well-formed local moments. In particular, we predict that for such materials, the performance of eDMFT is likely to be the best among the methods discussed here. We hope to expand our database to investigate the performance of these methods for a much wider range of materials.

\section{Conclusions}

In conclusion, we have proposed a new paradigm in which a wide range of DFT and beyond-DFT methods are applied to a selected set of target materials in a systematic and uniform manner in order to develop a systematic way of choosing the most accurate and efficient method for any given material. As a first demonstration, we have applied the GGA, GGA+U, mBJ, GW$_0$, B3LYP, and DFT+eDMFT methods to a set of four prototypical binary transition-metal oxides, MnO, NiO, FeO, and CoO, in the AFM phase, and evaluated their performance for the density of states, spectral functions and optical conductivity.  

For these materials, we find that eDMFT is a preferred methodology that can reasonably well reproduce the ARPES, PES/IPES, and optics experiments without any material-specific tuning of parameters.  
B3LYP also performs well in reproducing the main features of the DOS but has issues in describing the ARPES, and in some of the TMOs, the optics as well. 

 Although we have studied only four compounds here, we can predict that for moderately correlated AFM materials with open 3{\it d}-shells and well described local moments, the performance of eDMFT is likely to be superior among the methods discussed here. To establish similar conclusions for other materials classes, calculations are currently underway on a much broader range of materials. As we populate an open-source database containing the results, our findings show promise for accelerating the progress of computational materials discovery and design, especially as applied to correlated materials. Our work is thus representative of recent trends toward the integration of fundamental physical theories and computational methodologies with database-driven science and engineering.

\section{Methods}

In this work we use the full potential linear augmented plane wave (LAPW) method as described in the WIEN2k~\cite{WIEN2k} software for various DFT and beyond-DFT methods, such as the modified Becke-Johnson (mBJ) potential ~\cite{mbj} for meta-GGA, B3LYP ~\cite{doi:10.1063/1.464913,PhysRevB.74.155108} for hybrid functionals, all electron GW (FHI-gap software~\cite{FHI}) for Hedin's GW formalism, and embedded DMFT (eDMFT)~\cite{eDMFT2010,eDMFT2018} method for dynamical mean field theory. For mBJ and B3LYP we construct the initial wavefunction and eigenvalues with PBE functional in the generalized gradient approximation (GGA). For DFT, DFT+U, mBJ, B3LYP we use 20$\times$ 20 $\times$ 20 k-points and 0.01 Ry Gaussian broadening for computing DOS.

\textbf{GGA+U:} While the linear response theory tends to give smaller values of $U$ and usually quite accurately predicts the energetic and structural properties, the same $U$ is usually too small for a proper description of the spectrum~\cite{U1,U2,U3,PhysRevB.82.045108}. The constrained-DFT gives larger values of $U$, which are often too large when compared with the experiment. This is because DFT+U solves the impurity problem within the Hartree-Fock method, and hence all the higher order Feynman-diagrams (beyond the exchange) should be accounted for in the method which computes the effective $U$.  The U values in GGA+U are 6.04, 7.05, 5.91, 6.88 eV for MnO, NiO, FeO, and CoO respectively and obtained from Ref~\cite{PhysRevB.74.155108}.

\textbf{DFT+DMFT:} In eDMFT method~\cite{eDMFT2010,eDMFT2018} we use the LDA functional and the LAPW basis set as implemented in WIEN2k~\cite{WIEN2k}. The continuous time quantum Monte Carlo method~\cite{haule2007} is used to solve the quantum impurity problem that is embedded within the Dyson equation for the solid, to obtain the local self-energy for the TM $d$ orbitals. The self-energy is then analytically continued with the maximum entropy method from the imaginary to the real axis, continuing the local cumulant function, to obtain the partial density of states. In eDMFT, where all such higher-order Feynman diagrams are explicitly calculated by the impurity solver, the amount of screening by the degrees of freedom not included in the method is substantially reduced, and the values of $U$ are larger and are quite successfully predicted by the self-consistent constrained method. A fine k-point mesh of at least 10$\times$ 10 $\times$ 10 k-points in Monkhorst-Pack k-point grid and a total 100 million Monte Carlo steps for each iteration are used for the AFM phase of the TMO at T=300K. To avoid tuning parameters, the Coulomb interaction $U$ and Hund's coupling $J_{H}$  are fixed at 10.0 eV and 1.0 eV respectively for all four TMOs. These values are computed by the constrained-eDMFT method. We use exact double counting between LDA and DMFT~\cite{exact} and we also compare our results with the fully localized limit (FLL) double counting ~\cite{FLL} scheme. 

\textbf{GW:} We perform single-shot GW and GW$_0$ using FHI-gap software package~\cite{FHI} where GW self-energy is computed within the all-electron LAPW basis of WIEN2K. We use 4$\times$ 4 $\times$ 4 k-point grids and include unoccupied bands with energy up to 50 Ry. We also include high-energy local orbitals in the GW$_0$ calculations. About 1000 k-points are considered for computing the DOS, where 
we first compute the quasiparticle energies in a sparse k-mesh and then interpolate to a much finer k-mesh. The muffin tin radii (in Bohr) for Mn and O atoms are (2.10, 1.77) for MnO; (2.05, 1.75) for FeO; (1.97, 1.75) for CoO and NiO~\cite{PhysRevB.82.045108}. The Gaussian broadening and k-point sampling for computing DOS are kept at least $\sim$ 0.01 Ry and 10$\times$ 10 $\times$ 10, respectively. Similar values were used in the Ref.~\onlinecite{PhysRevB.82.045108}.

\textbf{Optical Absorption Computations:} To compute that within eDMFT, we obtain the imaginary part of the dielectric function from the real part of the optical conductivity and then perform the Kramers-Kronig (KK) operations to compute the absorption coefficients.

\section{Acknowledgment } 
We thank Steven Louie, Hong Jiang, and Zhenglu Li for helpful discussions related to GW calculations. We thank G. L. Pascut for helpful discussions. The computations were performed at the XSEDE, Rutgers HPC (RUPC). This research also used resources from the Rutgers Discovery Informatics Institute~\cite{RDI2}, which are supported by Rutgers and the State of New Jersey. This research was funded by NSF DMREF DMR-1629059 and NSF DMREF DMR-1629346.

\section{Data availability}
The data that support the findings of this study are available from the corresponding authors upon request. In the future, the data will be available in the form of an open-source database that is currently under construction.

\section{Contributions}
S.M carried out the calculations. All authors discussed the results and co-wrote the paper.

\section{References} 
\bibliography{SM-bib}

\end{document}